\pdfoutput=1
\documentclass[conference]{IEEEtran}
\IEEEoverridecommandlockouts

\usepackage[utf8]{inputenc} 
\usepackage[T1]{fontenc}    
\usepackage[hidelinks]{hyperref}  
\usepackage{url}            
\usepackage{booktabs}       
\usepackage{amsfonts}       
\usepackage{nicefrac}       
\usepackage{microtype}      
\usepackage[dvipsnames]{xcolor}  
\usepackage{xspace}
\usepackage{graphicx}
\usepackage{subcaption}
\usepackage{wrapfig}
\usepackage{soul} 
\usepackage{array}
\usepackage{ifthen}
\usepackage{siunitx}
\usepackage{balance}  
\usepackage{flushend}  

\usepackage[%
    backend=biber,
    style=numeric-comp,
    sorting=none,        
    sortcites=true,      
    block=none,          
    indexing=false,      
    citereset=none,      
    isbn=false,          
    url=true,            
    doi=false,           
    natbib=true,         
    maxnames=18,         
    minnames=12,         
    giveninits=true      
]{biblatex}
\newcolumntype{L}[1]{>{\raggedright\let\newline\\\arraybackslash\hspace{0pt}}m{#1}}
\newcolumntype{C}[1]{>{\centering\let\newline\\\arraybackslash\hspace{0pt}}m{#1}}
\newcolumntype{R}[1]{>{\raggedleft\let\newline\\\arraybackslash\hspace{0pt}}m{#1}}

\usepackage{listings}
\usepackage{xcolor}
\usepackage{float}
\newfloat{lstfloat}{htbp}{lop}
\floatname{lstfloat}{Listing}
\definecolor{commentsColor}{rgb}{0.497495, 0.497587, 0.497464}
\definecolor{keywordsColor}{rgb}{0.000000, 0.000000, 0.635294}
\definecolor{stringColor}{rgb}{0.558215, 0.000000, 0.135316}
\lstdefinestyle{mystyle}{
    basicstyle=\scriptsize\ttfamily,
    captionpos=b,
    breaklines=true,
    breakindent=0.5em,
    postbreak=\raisebox{0ex}[0ex][0ex]{\ensuremath{\color{red}\hookrightarrow}},  
    tabsize=2,
    frame=b,
    showstringspaces=false,
    numberstyle=\tiny\color{commentsColor},
    rulecolor=\color{black},
    commentstyle=\color{commentsColor}\textit,
    stringstyle=\color{stringColor},
    keywordstyle=\color{keywordsColor},
    emphstyle=\color{keywordsColor},
    escapeinside={(*@}{@*)},
}
\AtBeginDocument{\DeclareCaptionSubType{lstlisting}}

\title{Large Language Models for Compiler Optimization}
\makeatletter
\newcommand{\printfnsymbol}[1]{%
  \textsuperscript{\@fnsymbol{#1}}%
}
\makeatother

\author{%
\IEEEauthorblockN{Chris Cummins$^{\dagger*}$\thanks{$\dagger$ Core contributors. *Corresponding author: cummins@meta.com}, Volker Seeker$^\dagger$, Dejan Grubisic$^\dagger$,\\Mostafa Elhoushi, Baptiste Roziere, Jonas Gehring, Fabian Gloeckle,\\Kim Hazelwood, Gabriel Synnaeve, Hugh Leather$^\dagger$}
\IEEEauthorblockA{\emph{Meta AI}}
\and
\IEEEauthorblockN{Youwei Liang}
\IEEEauthorblockA{\emph{UC San Diego}}
}

\begin{document}

\maketitle

\begin{abstract}
We explore the novel application of Large Language Models to code optimization. We present a 7B-parameter transformer model trained from scratch to optimize LLVM assembly for code size. The model takes as input unoptimized assembly and outputs a list of compiler options to best optimize the program. Crucially, during training, we ask the model to predict the instruction counts before and after optimization, and the optimized code itself. These auxiliary learning tasks significantly improve the optimization performance of the model and improve the model’s depth of understanding.

We evaluate on a large suite of test programs. Our approach achieves a 3.0\% improvement in reducing instruction counts over the compiler, outperforming two state-of-the-art baselines that require thousands of compilations. Furthermore, the model shows surprisingly strong code reasoning abilities, generating compilable code 91\% of the time and perfectly emulating the output of the compiler 70\% of the time.
\end{abstract}

\section{Introduction}

There is increasing interest in Large Language Models (LLMs) for software engineering domains such as code generation~\cite{li_starcoder_2023, li_competition-level_2022, openai2023gpt4, allal2023santacoder, chowdhery2022palm, fried_incoder_2023, gunasekar2023textbooks, chen_evaluating_2021, llama-code}, code translation~\cite{transcoder, armengol2021learning, transcoder-ir}, and code testing~\cite{ye_2021, titanfuzz, schäfer2023adaptive}. Models such as Code Llama~\cite{llama-code}, Codex~\cite{chen_evaluating_2021}, and ChatGPT~\cite{chatgpt} have a good statistical understanding of code and suggest likely completions for unfinished code, making them useful for editing and creating software. However, it appears they have not been trained specifically to optimize code. ChatGPT, for instance, will make minor tweaks to a program such as tagging variables to be stored as registers, and will even attempt more substantial optimizations like vectorization, though it easily gets confused and makes mistakes, frequently resulting in incorrect code.

Prior works on machine learning-guided code optimization have used hand-built features~\cite{mlgo, wang2018machine, ml4sysreview}, all the way to graph neural networks (GNNs)~\cite{coreset, programl}. However, in all cases, the way the input program is represented to the machine learning algorithm is incomplete, losing some information along the way. For example, MLGO~\cite{mlgo} uses numeric features to provide hints for function inlining, but cannot faithfully reproduce the call graph or control flow, etc. PrograML~\cite{programl} forms graphs of the program to pass to a GNN, but it excludes the values for constants and some type information which prevents reproducing instructions with fidelity. 

In this work, we ask: can Large Language Models learn to optimize code? LLMs can accept source programs, as is, with a complete, lossless representation. Using text as the input and output representation for a machine learning optimizer has desirable properties: text is a universal, portable, and accessible interface, and unlike prior approaches is not specialized to any particular task.

We started our investigation into the code-optimizing power of LLMs by replicating the optimizing transformations present in compilers, targeting the industry standard LLVM~\cite{llvm} compiler. LLVM's optimizer is extremely complex and contains thousands of rules, algorithms, and heuristics in over 1M lines of C++ code. Our expectation was that while LLMs have shown great progress in natural language translation and code generation tasks, they would be incapable of emulating such a complex system. Understanding and applying compiler optimizations require multiple levels of reasoning, arithmetic computation capabilities, and applying complex data structure and graph algorithms, which are capabilities LLMs have shown to lack~\cite{asher2023limits,qian2022limitations}.

We thought this would be a paper about the obvious failings of LLMs that would serve as motivation for future clever ideas to overcome those failings. We were entirely taken by surprise to find that in many cases a sufficiently trained LLM can not only predict the best optimizations to apply to an input code, but it can also directly perform the optimizations without resorting to the compiler at all!

Our approach is simple. We begin with a 7B-parameter LLM architecture, taken from LLaMa 2~\cite{llama}, and initialize it from scratch. We then train it on millions of examples of LLVM assembly, coupled with the best compiler options found by a search for each assembly, as well as the resulting assembly from performing those optimizations. From these examples alone the model learns to optimize code with remarkable accuracy.

Our singular contribution is the first application of LLMs to optimizing code. We construct LLMs solely for the purpose of compiler optimization and show that they achieve a single-compile 3.0\% improvement in code size reduction over the compiler versus a search-based approach which achieves 5.0\% with $2.5e^9$ compilations and versus state of the state-of-the-art ML approaches that cause regressions and require thousands of compilations. We provide auxiliary experiments and code examples to further characterize the potential and limits of LLMs for code reasoning. Overall we find their efficacy remarkable and think that these results will be of interest to the community.
\section{Pass Ordering with LLMs}

\begin{figure*}
\begin{center}
\centerline{\includegraphics[width=1\textwidth]{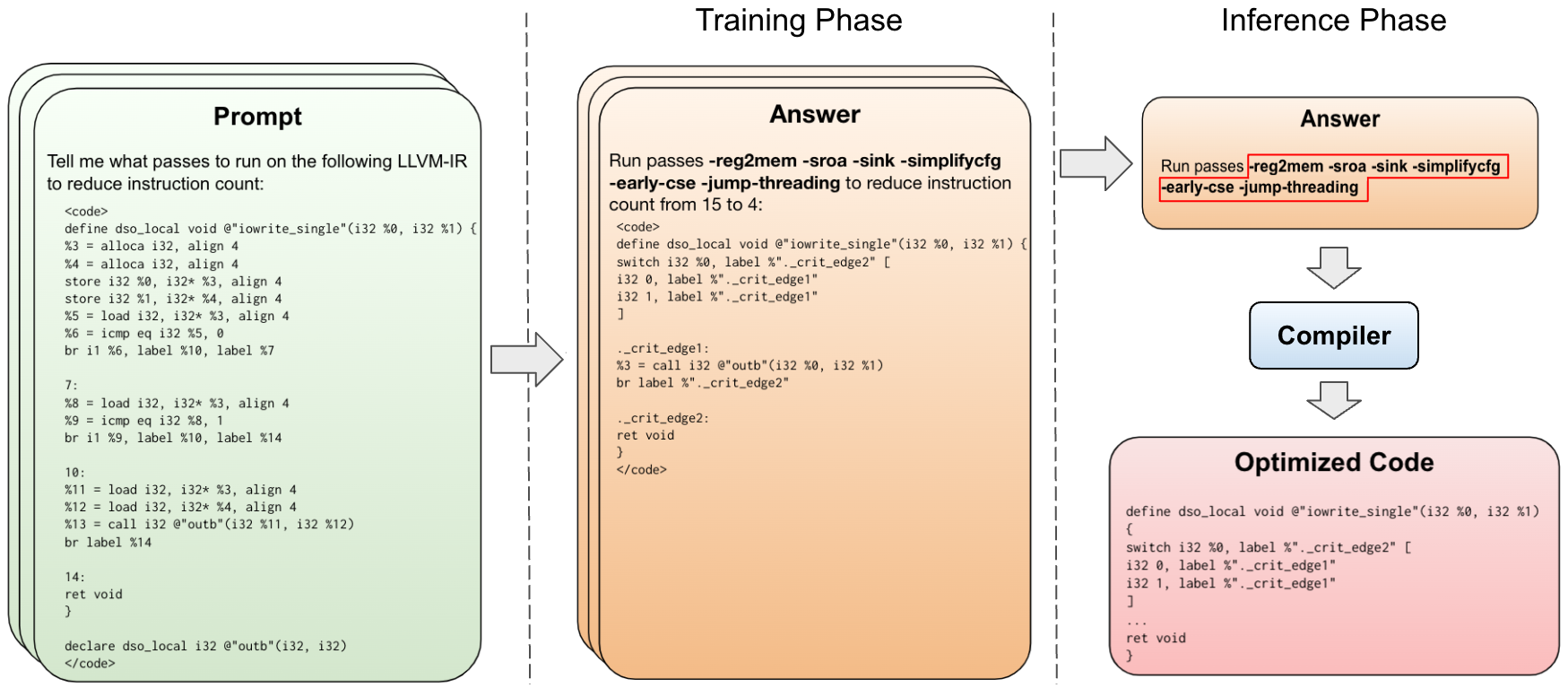}}
\caption{%
    Overview of our approach, showing the model input (Prompt) and output (Answer) during training and inference. The prompt contains unoptimized code. The answer contains an optimization pass list, instruction counts, and the optimized code. During inference we generate only the optimization pass list which we feed into the compiler, ensuring that the optimized code is correct.
}
\label{figures/overview}
\end{center}
\end{figure*}

In this work we target compiler pass ordering. The pass ordering task is to select from the set of optimizing transformation passes available in a compiler the list of passes that will produce the best result for a particular input code. Manipulating pass orders has been shown to have a considerable impact on both runtime performance and code size~\cite{fursin_evaluating_2005,ml4sysreview}.

Machine learning approaches to this task have shown good results previously, but struggle with generalizing across different programs~\cite{compilergym}. Previous works usually need to compile new programs tens or hundreds of times to try out different configurations and find out the best-performing option, making them impractical for real-world use. We hypothesized that a large language model with sufficient reasoning power would be able to learn to make good optimization decisions without needing this.

\begin{table}
  \caption{%
    Training data. Each LLVM-IR function is autotuned and used to create a (Prompt, Answer) pair. The $n$ tokens column shows the number of tokens when the prompt is encoded using the Llama 2~\cite{llama} tokenizer.
}
  \label{tables/training-data}
  \centering
  \begin{tabular}{lC{1.13cm}C{1.5cm}C{1.2cm}c}
    \toprule
    & $n$ functions & unoptimized instruction count & size on disk & $n$ tokens \\
    \midrule
    Handwritten & 610,610 & 8,417,799 & 653.5 MB & 214,746,711 \\
    Synthetic & 389,390 & 13,775,149 & 352.3 MB & 158,435,151 \\
    \midrule
    Total & 1,000,000 & 16,411,249 & 1.0 GB & 373,181,862 \\
    \bottomrule
  \end{tabular}
\end{table}

Most prior work on LLMs for code operates on source languages such as Python. Instead, for the pass ordering problem we require reasoning at the lower level of compiler assembly, known as the Intermediate Representation (IR). While there exist curated datasets of source languages for pretraining LLMs (e.g.~\cite{the-stack,the-pile,codesearch-net}), compiler IRs do not make up a significant portion of these datasets, and though models like ChatGPT show some promise of understanding, their ability to reason about IR is far inferior to source languages.

We target optimizing LLVM pass orders for code size as in prior works~\cite{mlgo,compilergym}, using IR instruction count as an (imperfect) proxy for binary size. The approach is agnostic to the chosen compiler and optimization metric, and we intend to target runtime performance in the future. For now, optimizing for code size simplifies the collection of training data.

\begin{table}
  \caption{%
    Test data.
  }
  \label{tables/test-data}
  \centering
  \begin{tabular}{lC{1.25cm}C{1.75cm}C{1.6cm}}
    \toprule
    & $n$ functions & unoptimized instruction count & -Oz instruction count \\
    \midrule
    AI-SOCO~\cite{ai-soco} & 8,929 & 97,800 & 47,578 \\
    ExeBench~\cite{exebench} & 26,806 & 386,878 & 181,277 \\
    POJ-104~\cite{poj104} & 310 & 8,912 & 4,492 \\
    Transcoder~\cite{transcoder-ir} & 17,392 & 289,689 & 129,611 \\
    CSmith~\cite{csmith} & 33,794 & 647,815 & 138,276 \\
    YARPGen~\cite{yarpgen} & 12,769 & 285,360 & 144,539 \\
    \midrule
    Total & 100,000 & 1,716,354 & 645,773 \\
    \bottomrule
  \end{tabular}
\end{table}

\subsection{Prompts}

We present the model with an unoptimized LLVM-IR (such as emitted by the \emph{clang} frontend) and ask it to produce a list of optimization passes that should be applied to it. Figure~\ref{figures/overview} shows the format of the input prompt and output text.

In this work, we target LLVM 10 and use the optimization flags from \texttt{opt}. There are 122 optimization passes to choose from and passes can be selected more than once in a single sequence. We also include the 6 meta-flags (-O0, -O1, -O2, -O3, -Oz, and -Os) that may each occur only once per pass list. Pass lists can be any length, though in our experiments we found typically up to 9 passes long, for a combinatorial search space of around $10^{18}$.

As shown in Figure~\ref{figures/overview}, we also include two auxiliary tasks: i) generating the instruction counts of the code before and after the optimizations are applied and ii) generating the output IR after the optimizations are applied. We hypothesize that these would enable better pass-ordering decisions by forcing a deep understanding of the mechanics of code optimization. We verify this experimentally in Section~\ref{section/codegen-ablation}.

While the model is trained to generate instruction counts and optimized IR, we do not need those auxiliary tasks for deployment. All we need to do is generate the pass list which we then execute using the compiler. We thus sidestep the problems of correctness that plague techniques that require the output of the model to be trustworthy~\cite{armengol2023slade,transcoder-ir,armengol2021learning,transcoder}.

\subsection{LLVM-IR Normalization}

We normalize the LLVM-IR that is used for training the LLM using the following rules: we discard comments, debug metadata and attributes, and ensure consistent whitespace by feeding the IR through a custom lexer that retains newlines but standardizes other whitespace and strips indentation. We do this to reduce the length of the LLVM-IR to make maximum use of the limited input size of the LLM (Section~\ref{section/model-architecture}). The code in Figure~\ref{figures/overview} has been processed in this manner.
\section{The Model}
\label{section/model}

We use the ubiquitous transformer architecture~\cite{transformer}. The transformer is an artificial neural network that employs self-attention over a fixed-size context window.

The input text is first tokenized into words and subword units. These are embedded into continuous vector representations and provided as input to the transformer's encoder, where self-attention mechanisms capture contextual relationships between tokens to encourage the model to understand and process the input text's semantic structure.

The output text is produced by iteratively generating one token at a time. The decoder takes the encoded input along with any previously generated tokens and uses self-attention to predict the next token in the sequence. We greedily sample during decoding to select the most likely token sequence. This process continues until an end-of-sequence token is generated or a predefined maximum length is reached.

\subsection{Model Architecture}%
\label{section/model-architecture}

We use the same model architecture and Byte Pair Encoding (BPE)~\cite{bpe} tokenizer as Llama 2~\cite{llama}, but train our model from scratch. We use the smallest of the Llama 2 configurations: 32 attention heads, 4,096 hidden dimensions, and 32 layers, for a total of 7B parameters.

The maximum length of a (prompt, answer) pair is defined by the sequence length. In this work, we use a sequence length of 2,048 tokens. The Llama 2 tokenizer achieves an average of 2.02 characters per token when encoding LLVM-IR, so this provides an approximate upper limit on the longest LLVM-IR we can train on at 2KB (since 2KB prompt and 2KB answer $\approx$ 2,048 tokens).

\begin{figure}
    \centering
    \begin{subfigure}{\columnwidth}
        \centering
        \includegraphics[width=\columnwidth]{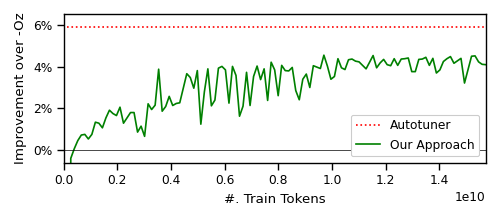}
        \caption{
            Performance of generated pass lists. 
        }
        \label{figures/perf-metrics}
    \end{subfigure}
    \begin{subfigure}{\columnwidth}
        \centering
        \includegraphics[width=\columnwidth]{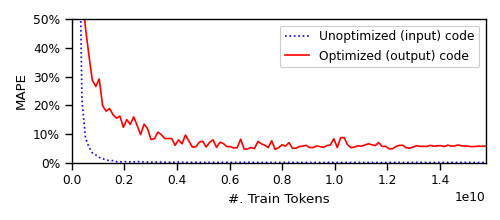}
        \caption{
            Accuracy at predicting instruction counts.
        }
        \label{figures/instcount-metrics}
    \end{subfigure}
    \begin{subfigure}{\columnwidth}
        \centering
        \includegraphics[width=\columnwidth]{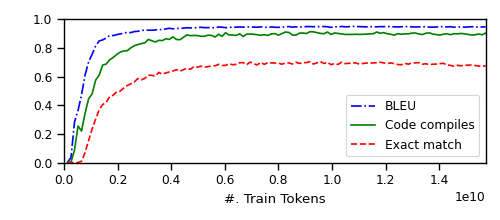}
        \caption{
            Model-optimized code metrics.
        }
        \label{figures/codegen-metrics}
    \end{subfigure}
    \caption{Performance on holdout validation set during training. We evaluate performance every 250 training steps (131M train tokens). Parity with -Oz is reached at 393M tokens and peak performance at 10.9B tokens.}
    \label{figures/validation-metrics}
\end{figure}

\subsection{Training Data}%
\label{section/training-data}

\begin{table*}
  \caption{%
    Performance of different approaches to pass ordering on a test set of unseen LLVM-IR functions from Table~\ref{tables/test-data}.  All metrics are \emph{w.r.t.\ }-Oz. \emph{Instructions saved} is summed over \emph{functions improved} and \emph{instructions regressed} is summed over \emph{functions regressed}. \emph{Overall improvement} is the sum total instruction count savings \emph{w.r.t\ }-Oz. The Autotuner achieves the best performance but requires 2.5B additional compilations (949 CPU-days). Our approach achieves 60\% of the gains of the autotuner without invoking the compiler once.
  }
  \label{tables/pass-ordering-results}
  \centering
  \begin{tabular}{lC{1.7cm}C{1.1cm}C{1.1cm}C{1.5cm}C{1.5cm}C{1.7cm}}
    \toprule
    & additional compilations & functions improved & functions regressed & instructions saved & instructions regressed & overall improvement \\
    \midrule
    Autotuner & 2,522,253,069 & 6,764 & 0 & 30,948 & 0 & 5.03\% \\
    AutoPhase~\cite{autophase} & 4,500,000 & 1,558 & 8,400 & 6,522 & 32,357 & -3.85\% \\
    Coreset-NVP~\cite{coreset} & 442,747 & 3,985 & 6,072 & 16,064 & 28,405 & -1.88\% \\
    Our Approach & 0 & 4,136 & 526 & 21,935 & 3,095 & 3.01\% \\
    \bottomrule
  \end{tabular}
\end{table*}

\begin{table}
  \caption{%
    Extending the models in Table~\ref{tables/pass-ordering-results} with ``-Oz backup''. If a model predicts a pass list \emph{other than} -Oz, it also evaluates -Oz and selects the best. This prevents regressions \emph{w.r.t\ }-Oz at the expense of additional compilations.
  }
  \label{tables/pass-ordering-results-with-backtracking}
  \centering
  \begin{tabular}{lcc}
    \toprule
    & additional compilations & overall improvement \\
    \midrule
    AutoPhase~\cite{autophase} & 4,600,000 & 1.02\% \\
    Coreset-NVP~\cite{coreset} & 542,747 & 2.55\% \\
    Our Approach & 5,721 & 3.52\% \\
    \bottomrule
  \end{tabular}
\end{table}

We assembled a large corpus of unoptimized LLVM-IR functions, summarized in Table~\ref{tables/training-data}. We extracted the functions from datasets of publicly available handwritten C/C++ code and supplemented this with synthetic code generated by C/C++ compiler test generators. In total, our training corpus comprises 1,000,000 deduplicated IR functions, totaling 373M training tokens. We operate at the level of individual IR functions rather than entire modules to maximize the amount of data we can fit inside a 2,048-token sequence length.

To find the list of optimization passes that will produce the smallest instruction count we employ \emph{autotuning}. Our autotuner combines random search and all-to-all results broadcasting between functions, inspired by the work of Liang et.\ al.~\cite{coreset}. For each function we run random search for a fixed amount of time (780 seconds) and then minimize the best pass list by iteratively removing individual randomly chosen passes to see if they contribute to the instruction count. If not, they are discarded. After performing this on each of the functions we aggregate the set of unique best pass lists and broadcast them across all other functions. Thus, if a pass list was found to work well on one function it is tried on all others.

In total, the autotuner compiled each training program an average of 37,424 times, achieving a 5.8\% improvement in instruction count reduction over the baseline fixed pass ordering in the compiler provided by -Oz. For our purposes, this autotuning serves as a gold standard for the optimization of each function. While the instruction count savings discovered by the autotuner are significant, the computational cost to reach these wins was 9,016 CPU days. The goal of this work is to achieve some fraction of the performance of the autotuner using a predictive model that does not require running the compiler thousands of times.

\subsection{Training}

Starting from randomly initialized weights, we trained the model for 30,000 steps on 64 V100s for a total training time of 620 GPU days. We use the AdamW optimizer~\cite{adamw} with $\beta_1$ and $\beta_2$ values of 0.9 and 0.95. We use a cosine learning rate schedule with 1,000 warm-up steps, a peak learning rate of $1e{-5}$, and a final learning rate of 1/10th of the peak. We used a batch size of 256 and each batch contains 524,288 tokens for a total of 15.7B training tokens. The full 30,000 steps of training is 7.7 epochs (iterations over the training corpus).

During training, we evaluated the model on a holdout validation set of 1,000 unseen IRs that were processed in the same manner as the training set. We evaluate every 250 steps.

\section{Evaluation}

In this section, we evaluate the ability of the model to generate pass lists for unseen code and to correctly perform optimization.

\subsection{Training Results}

Figure~\ref{figures/validation-metrics} shows the performance during training when evaluated on a holdout validation set of 1,000 unseen LLVM-IR functions. Peak validation performance was achieved by the model at 10.9B training tokens.

At peak performance, the code optimized using model-generated pass sequences contains 4.4\% fewer instructions than when optimized using the compiler's built-in pass ordering (-Oz). The autotuner achieves a greater instruction count reduction of 5.6\%, but this required 27 million compilations of the validation set. The model makes its predictions without invoking the compiler once. 

Figure~\ref{figures/instcount-metrics} shows the error of predicted input and output instruction counts. Prediction of instruction counts for unoptimized code rapidly approaches near-perfect accuracy. Prediction of output instruction count proves more challenging, reaching a Mean Average Percentage Error (MAPE) of 5.9\%.

Figure~\ref{figures/codegen-metrics} evaluates the quality of the generated code using three metrics. The \emph{BLEU}~\cite{bleu} score shows the similarity between the model-generated code and a reference ground-truth code produced by the compiler using the generated pass list. \emph{Code compiles} is the frequency that model-generated code compiles without error. \emph{Exact match} tracks the frequency that the model-generated code is a character-by-character match of the compiler-generated code when optimized using the generated pass list (i.e. how many times BLEU=1).

At peak performance, the model achieves an impressive 90.5\% rate of generating code that compiles without errors. Furthermore, a BLEU score of 0.952 shows that the model-optimized code closely approximates that of the compiler, and the exact match frequency is 70\%. For comparison, a baseline that simply copies the unoptimized code to the output would achieve a BLEU score of 0.531 and an exact match frequency of 0\%, demonstrating that significant manipulation of the input code is required to achieve such high scores.

By the end of training performance on the validation set had plateaued. We use the best-performing checkpoint and switch to a $100\times$ larger-scale evaluation for the remainder of the evaluation.

\begin{figure*}
    \begin{center}
    \centerline{\includegraphics[width=1\textwidth]{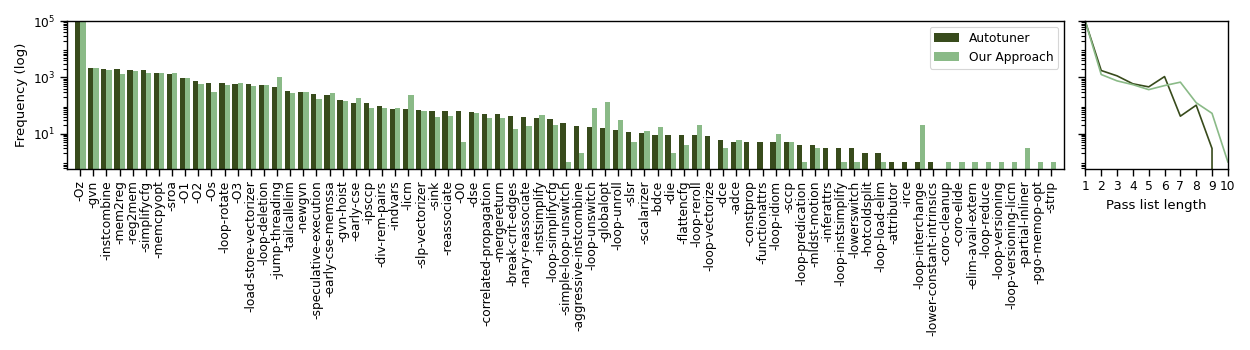}}
    \caption{%
        Frequency that passes occur in the pass list for each of the 100,000 test programs (left), and the length of pass lists (right). -Oz is the starting point for the autotuner and is the dominant result, being the best-found result for 93.2\% of autotuned test programs and appearing in an additional 0.6\% of pass lists as part of a longer sequence. The model-generated pass distribution tracks the autotuner but slightly overpredicts -Oz (94.3\%) and includes 9 passes that the autotuner used on the training set but not on the test set. Results are ordered by decreasing autotuner frequency.
        \vspace{-1.5em}
    }
    \label{figures/pass-frequency}
    \end{center}
\end{figure*}

\begin{figure*}
\setcaptiontype{lstlisting}
\begin{minipage}[t]{.27\textwidth}
    \begin{lstlisting}[language={LLVM},escapechar=ä]
define i32 @f1(i8 %0) {
 %2 = alloca i32, align 4
 %3 = alloca i8, align 1
 store i8 %0, i8* %3, align 1
 %4 = load i8, i8* %3, align 1
 %5 = zext i8 %4 to i32
 %6 = icmp sge i32 %5, 65
 br i1 %6, label %7, label %15

7:
 %8 = load i8, i8* %3, align 1
 %9 = zext i8 %8 to i32
 %10 = icmp sle i32 %9, 90
 br i1 %10, label %11, label %15

11:
 %12 = load i8, i8* %3, align 1
 %13 = zext i8 %12 to i32

ä\textcolor{lightgray}{\textit{\textbf{<snip 21 lines\ldots>}}}ä

33:
 %34 = load i32, i32* %2, align 4
 ret i32 %34
}
    \end{lstlisting}
    \subcaption{Input code (39 instructions).}
\end{minipage}
\hfill
\begin{minipage}[t]{.34\textwidth}
    \begin{lstlisting}[language={LLVM},escapechar=ä]
define i32 @f1(i8 %0) {
 %2 = zext i8 %0 to i32
 %.off = add i8 %0, 191
 %3 = icmp ult i8 %.off, 26
 br i1 %3, label %4, label %6

4:
 %5 = add nsw i32 %2, 191
 br label %10

6:
 %.reload16.off = add nsw i32 %2, 159
 %7 = icmp ult i32 %.reload16.off, 26
 br i1 %7, label %10, label %8

8:
 %9 = icmp eq i8 %0, 32
 %. = select i1 %9, i32 26, i32 1
 br label %10

10:
 %.0.reg2mem.0 = phi i32 [%5, %4], 
  [%., %8], [%.reload16.off, %6]
 ret i32 %.0.reg2mem.0
}
    \end{lstlisting}
    \subcaption{Autotuned code (14 instructions) using passes: {\scriptsize \texttt{-reg2mem -instcombine -Os -O1}}.}
\end{minipage}
\hfill
\begin{minipage}[t]{.33\textwidth}
    \begin{lstlisting}[language={LLVM},escapechar=ä]
define i32 @f1(i8 %0) {
 %2 = zext i8 %0 to i32
 %.off = add i8 %0, 191
 %3 = icmp ult i8 %.off, 26
 br i1 %3, label %6, label %._crit_edge

._crit_edge:
 %.off24 = add i8 %0, 159
 %4 = icmp ult i8 %.off24, 26
 br i1 %4, label %6, label %._crit_edge9

._crit_edge9:
 %5 = icmp eq i8 %0, 32
 %spec.select = select i1 %5, 
  i32 26, i32 1
 ret i32 %spec.select

6:
 %.sink = phi i32 [191, %1], 
  [159, %._crit_edge]
 %7 = add nsw i32 %.sink, %2
 ret i32 %7
}
    \end{lstlisting}
    \subcaption{Model-optimized code (13 instructions) and pass list: {\scriptsize \texttt{-reg2mem -simplifycfg -mem2reg -jump-threading -Os}}.}
\end{minipage}
\caption{%
    An example IR function where the model suggests a better pass list than the autotuner, despite having never seen this code before. For this function the autotuner tried 26k different pass orderings. The pass list generated by the model appears 5 times in the training set of 1,000,000 examples.%
}
\label{listings/better-than-autotuner}
\end{figure*}

\subsection{Comparison to State-of-the-Art}

In this experiment, we perform a large-scale evaluation of the LLM's ability to predict pass lists in comparison to baselines.

\textbf{Datasets} We aggregate a broad suite of benchmark datasets for evaluation, summarized in Table~\ref{tables/test-data}. We deduplicate and exclude IR functions identical to those we trained on. Our test data comprises code from a variety of domains including coding competitions (AI-SOCO~\cite{ai-soco}, POJ-104~\cite{poj104}), compiler test case generators (CSmith~\cite{csmith}, YARPGen~\cite{yarpgen}), and miscellaneous publicly available code (ExeBench~\cite{exebench}, Transcoder~\cite{transcoder-ir}).

\textbf{Baselines} We compare our approach to three baselines: AutoPhase~\cite{autophase}, Coreset-NVP~\cite{coreset}, and the Autotuner.

AutoPhase~\cite{autophase} is a reinforcement learning approach in which an agent is trained using Proximal Policy Optimization~\cite{ppo} to select the sequence of optimization passes that will maximize cumulative instruction count savings over a fixed-length episode. At each step, the program being optimized is represented to the agent as a 56-dimensional vector of instruction counts and other properties. We replicate the environment of~\cite{autophase} but use the implementation and expanded training regime from~\cite{compilergym} in which the agent is trained for 100,000 episodes. We train the agent on the same data as our language model (Table~\ref{tables/training-data}) and evaluate agent performance periodically during training on a holdout validation set. As in prior works, we use an action space and episode length of 45.

Coreset-NVP~\cite{coreset} is a technique that combines iterative search with a learned cost model. First, a greedy search is run on 17,500 benchmarks to determine a \emph{Core set} of best pass lists. Then a \emph{Neural Value Prediction} (NVP) is trained on the results of this search, using ProGraML~\cite{programl} graphs processed by a Graph Convolutional Network as program representation. At inference, Coreset-NVP predicts the normalized reward and tries the first few pass sequences with the highest normalized reward. The total number of passes it is allowed to try for each benchmark is 45, following prior works. We use author-provided model weights to perform inference on our test set.

Finally, we compare it to the Autotuner that we used to generate training data. We autotuned the test dataset in the same manner as the training data, described in Section~\ref{section/training-data}.

\textbf{Results} Table~\ref{tables/pass-ordering-results} summarizes the results. Our approach outperforms -Oz, AutoPhase, and Coreset-NVP across all datasets. Overall, the thousands of optimization attempts that are afforded to the autotuner enable it to discover the best-performing pass lists.

AutoPhase and Coreset-NVP are both able to identify pass lists that outperform -Oz but have an overall net negative impact on instruction count due to a large number of regressions. We propose a simple ``-Oz backup'' extension to overcome this: if a model predicts a pass list \emph{other than} -Oz, we also run -Oz and select the best of the two options. This prevents regressions \emph{w.r.t.\ }-Oz, but increases the number of additional compilations by the number of times the model predicts a pass list other than -Oz. Table~\ref{tables/pass-ordering-results-with-backtracking} shows the results of the techniques when evaluated in this manner. While this does not help the models find further improvements, the lack of regressions means that AutoPhase and Coreset-NVP now achieve overall improvements over -Oz, though still less than the LLM with or without the -Oz backup.

\begin{figure}
\begin{center}
\centerline{\includegraphics[width=1\columnwidth]{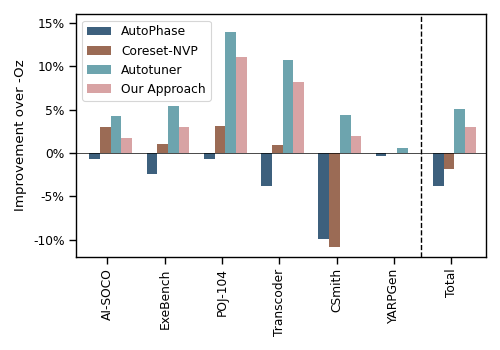}}
\caption{%
    Improvement over -Oz by dataset. Handwritten code optimizes more.
}
\label{figures/improvement-by-dataset}
\end{center}
\end{figure}
\begin{figure}
\begin{center}
\centerline{\includegraphics[width=1\columnwidth]{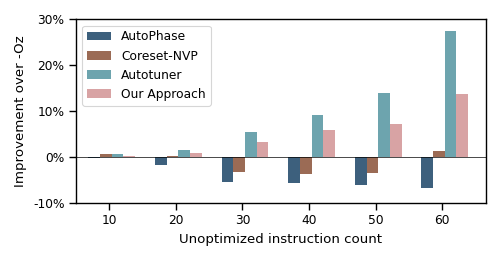}}
\caption{%
    Improvement over -Oz by input size. Larger codes optimize more.
}
\label{figures/improvement-by-size}
\end{center}
\end{figure}

\subsection{Evaluation of Generated Pass Lists}

Figure~\ref{figures/pass-frequency} shows the frequency with which passes are selected by the autotuner and our model from the previous experiment. The distribution of passes selected by the model broadly tracks the autotuner. -Oz is the most frequently optimal pass. Excluding -Oz, model-generated pass lists have an average length of 3.4 (max 10), and autotuner pass lists have an average length of 3.1 (max 9). 105 of the pass lists generated by the model never appear in the training data.

In 710 cases the model-generated pass lists outperform the autotuner on the test set, though improvements are typically small. Listing~\ref{listings/better-than-autotuner} shows an example where the model-generated pass list simplifies control flow to fewer blocks, saving one further instruction.

Figure~\ref{figures/improvement-by-dataset} breaks down the improvement of each approach to pass ordering by benchmark dataset. The biggest improvements over -Oz is found in the POJ-104 and Transcoder datasets, which both aggregate large amounts of handwritten code, while YARPGen, a random program generator for testing compilers, has the fewest opportunities for improving over -Oz.

We discovered that there is a strong correlation between the input program size and the potential performance improvement over -Oz that is found by both the autotuner and the model. Figure~\ref{figures/improvement-by-size} plots this trend, showing clearly that larger programs have more opportunities to improve over -Oz.

\subsection{Evaluation of Generated Code}

%

In this section, we evaluate the quality of model-generated code. To do this we ran the auxiliary training task of generating optimized code for all 100k functions in the test set. Note that this is not required to generate the pass lists evaluated in the previous section. We have made minor edits to the code samples in this section for brevity such as omitting superfluous statements and shortening identifier names.

In 90.3\% of cases, the model-generated optimized IR compiles, and in 68.4\% of cases the output IR matches character-for-character the ground truth generated by the compiler. We taxonomize the different classes of errors for the 9.7\% of cases where the generated IR does not compile in Table~\ref{tables/compile-errors}, and Listing~\ref{listings/compile-errors} provides code examples.

\begin{table}
  \caption{%
    Compiler errors of model-optimized code on 100,000 unseen inputs.
}
  \label{tables/compile-errors}
  \centering
  \begin{tabular}{lc}
    \toprule
    error category & $n$ \\
    \midrule
type error & 5,777 \\
instruction forward referenced & 1,521 \\
undefined value & 1,113 \\
invalid redefinition & 616 \\
syntax error & 280 \\
invalid value for constant & 144 \\
undefined function & 112 \\
index error & 98 \\
other & 83 \\
\midrule
Total & 9,744 \\
    \bottomrule
  \end{tabular}
\end{table}

\begin{figure}
\setcaptiontype{lstlisting}
\begin{minipage}[t]{1\columnwidth}
    \begin{lstlisting}[language={LLVM},escapechar=ä]
äerror: '\%15' \textbf{defined with type 'i32' but expected 'i1'}ä
%or.cond = or i1 %14, %15
    \end{lstlisting}
    \subcaption{The model defined \lstinline{\%15} as an integer but later tried to use it as a bool \emph{(type error)}.}
\end{minipage}
\begin{minipage}[t]{1\columnwidth}
    \begin{lstlisting}[language={LLVM},escapechar=ä]
äerror: constant expression type mismatchä
@.str = private unnamed_addr constant [ä\textbf{493 x i8}ä]
    c"ä\textcolor{lightgray}{\textbf{\textit{<snip 492 chars \ldots>}}}ä", align 1
    \end{lstlisting}
    \subcaption{The model omitted a single character when transcribing a 493-character string-literal from the input code \emph{(type error)}.}
\end{minipage}
\begin{minipage}[t]{1\columnwidth}
    \begin{lstlisting}[language={LLVM},escapechar=ä]
äerror: floating point constant invalid for typeä
%1 = tail call i32 @f1(float ä\textbf{-0.47799998483256463}ä, 
                       float ä\textbf{-1.8159999847412109}ä)
    \end{lstlisting}
    \subcaption{LLVM requires exact decimal values for floating-point constants. These model-generated values have repeating decimals in binary so are rejected \emph{(invalid value for constant)}.}
\end{minipage}
\caption{Compiler errors in model-optimized code.}
\label{listings/compile-errors}
\end{figure}

\begin{figure}
\setcaptiontype{lstlisting}
\begin{minipage}[t]{1\columnwidth}
    \begin{lstlisting}[language={LLVM},escapechar=ä]
define hidden signext i8 @f1() #0 {
  %1 = alloca i64, align 8
  store ä\textbf{i64 3718042838174166437}ä, i64* %1, align 8
  %2 = load i64, i64* %1, align 8
  %3 = ä\textbf{trunc i64 \%2 to i8}ä
  ret i8 %3
}
    \end{lstlisting}
    \subcaption{Input unoptimized code.}
\end{minipage}
\begin{minipage}[t]{.48\columnwidth}
    \begin{lstlisting}[language={LLVM},escapechar=ä]
define hidden signext i8
@f1() #0 {
  ret i8 ä\textbf{165}ä
}
    \end{lstlisting}
    \subcaption{Desired optimized code.}
\end{minipage}
\hfill
\begin{minipage}[t]{.48\columnwidth}
    \begin{lstlisting}[language={LLVM},escapechar=ä]
define hidden signext i8 
@f1() #0 {
  ret i8 ä\textbf{1}ä
}
    \end{lstlisting}
    \subcaption{Model-generated code.}
\end{minipage}
\caption{An example where the model generates compilable code but fails to compute the correct answer for a numeric expression. Producing the correct result for this expression requires non-trivial mathematical reasoning.}
\label{listings/hard-math}
\end{figure}

%
%
%
\begin{figure}
\begin{lstlisting}[language={LLVM},escapechar={ä},label={listings/right-code-wrong-pass-list},caption={%
    An example where the model generates correctly optimized code but fails to produce the pass list needed to produce the desired code. The model-optimized code and instruction count predictions match the performance of the autotuner, but the model omitted the \texttt{-mem2reg} pass needed to achieve this code. The model-generated pass list yields~10 instructions instead of~7.%
}]
ä\textit{Run passes \textbf{-instcombine -simplifycfg} to reduce}ä
ä\textit{instruction count from \textbf{14} to \textbf{7}:}ä

define dso_local i32 @f1(i32 %0) {
 %2 = load i64, i64* getelementptr inbounds(
   %struct.t2, %struct.t2* @gvar, i64 0, i32 0), align 8
 %3 = icmp eq i64 %2, 0
 %4 = icmp eq i32 %0, 0
 %or.cond = or i1 %3, %4
 %5 = load i32, i32* @S64_MAX, align 4
 %6 = select i1 %or.cond, i32 %5, i32 %0
 ret i32 %6
}
\end{lstlisting}
\end{figure}

\begin{figure}
\setcaptiontype{lstlisting}
\begin{minipage}{.52\columnwidth}
    \begin{lstlisting}[language={LLVM},escapechar=ä]
define i32 @f1(
  i32 %0, 
  i32 %1
) align 2 {
 br label %3

3:
 %i = phi i32 [%7, %6], [2, %2]
 %4 = mul nsw i32 %i, %i
 %5 = icmp sgt i32 %4, %1
 br i1 %5, label %8, label %6

6:
 %7 = add i32 %i, 1
 br label %3

8:
 ä\textbf{ret i32 2}ä
}
    \end{lstlisting}
    \subcaption{Desired optimized code.}
    \label{listings/inf-loop-autotuner}
\end{minipage}%
\hfill
\begin{minipage}{.44\columnwidth}
    \begin{minipage}{\columnwidth}
        \begin{lstlisting}[language={C},escapechar=ä]
int f1(int x, int y) {
  int i = 2;
  while (i * i < y) {
    i += 1;
  }
  return 2;
}
        \end{lstlisting}
        \subcaption{Equivalent (hand-written) C code.}
    \end{minipage}
    \begin{minipage}{\columnwidth}
        \begin{lstlisting}[language={LLVM},escapechar=ä]
define i32 @f1(
  i32 %0, 
  i32 %1
) align 2 {
 ä\textbf{ret i32 2}ä
}
        \end{lstlisting}
        \subcaption{Model-optimized code.}
        \label{listings/inf-loop-model}
    \end{minipage}
\end{minipage}
\caption{%
    An example of an unsafe optimization by the model. The 33-instruction input program (not shown) contains a loop that is not always safe to optimize away. For example, when \lstinline{y = INT_MAX} the loop never terminates.
}
\label{listings/inf-loop}
\end{figure}

\begin{figure}
\begin{center}
\centerline{\includegraphics[width=1\columnwidth]{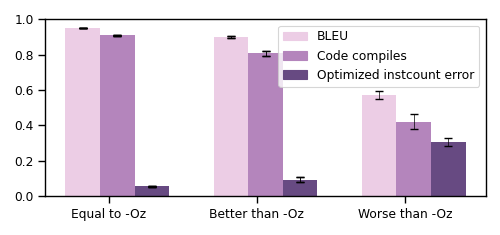}}
\caption{%
    Model-optimized code quality as a function of the performance of the generated pass list. Code quality is lower when the pass list performs worse than -Oz. The model-optimized code resembles the ground truth less (lower BLEU score), the code is less likely to compile, and the model struggles to estimate the instruction count (higher error). Error bars show 95\% confidence intervals.
}
\label{figures/accuracy-by-perf}
\end{center}
\end{figure}

Most challenging to evaluate are the 21.9\% of cases where the model-optimized code compiles but is not a character-by-character match with the compiler. There are two challenges: the first is that text precision metrics such as BLEU score are sensitive to differences in the code such as variable names and commutative operand order that do not affect the behavior of the code. Tools like LLVM-Canon~\cite{llvm-canon} can help here but come with their own set of drawbacks. However, in many cases, it is unclear whether the behavior of two IRs is the same, so the second challenge we face is in evaluating \emph{semantic} equivalency. Since not all of the datasets we use for testing provide driver scripts and input datasets for their code, we cannot use execution-based equivalence checks such as differential testing~\cite{difftest}.

\begin{figure*}
    \centering
    \includegraphics[width=\textwidth]{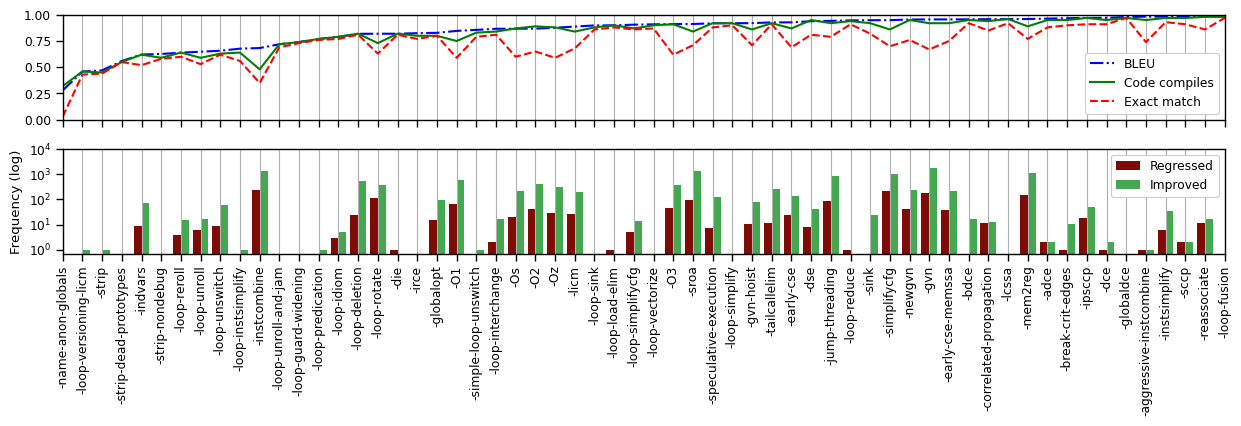}
    \caption{%
        Training a model to predict single optimization passes. The top subplot evaluates the quality the of generated code for the corresponding pass (ordered by BLEU score). The bottom subplot shows the frequency that the corresponding pass contributed to an improvement or regression of instruction count over -Oz.
    }
    \label{figures/single-pass-translation}
\end{figure*}

\begin{figure}
\begin{center}
\centerline{\includegraphics[width=1\columnwidth]{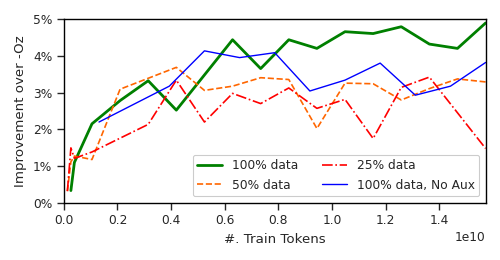}}
\caption{%
    Ablating the impact of training data size and the auxiliary co-training task of generating optimized code (denoted \textit{No Aux}). Data size is measured as a number of training examples. The graph shows performance on a holdout validation set during training.
}
\label{figures/ablations}
\end{center}
\end{figure}
\begin{table}
  \centering
  \caption{%
    Ablation experiments. We evaluate the impact of varying training data size and of training the model to generate the optimized code. We train each model for 30k steps and report performance of the best model checkpoint on a holdout validation set of 1,000 unseen IR functions.
  }
  \label{tables/ablations}
  \footnotesize
  \begin{tabular}{C{2cm}C{2cm}C{2cm}}
    \toprule
    $n$ training examples & generate optimized code? & overall improvement \\
    \midrule
    1,000,000 & \checkmark & 4.95\% (---) \\
    500,000 & \checkmark & 3.91\% (-21\%) \\
    250,000 & \checkmark & 3.74\% (-24\%) \\
    \midrule
    1,000,000 & $\times$ & 4.15\% (-16\%) \\
    \bottomrule
  \end{tabular}
\end{table}

Listing~\ref{listings/hard-math} shows an example of model-generated code that has incorrect program semantics. Here, the lower 8 bits of a 64-bit literal are truncated and returned. The compiler performs this calculation and substitutes the correct value. The model recognizes that the expression can be calculated at compile time but fails to compute the correct value. This type of mathematical reasoning is a known weakness of LLMs~\cite{qian2022limitations}.

Sometimes the model generates correctly-optimized code but fails to produce the pass list needed to achieve it. Listing~\ref{listings/right-code-wrong-pass-list} shows one such example. A further class of error is when the model makes unsafe optimizations by failing to analyze the input code. Listing~\ref{listings/inf-loop} shows an example.

We observe an interesting connection between the quality of pass lists and the corresponding optimized code, shown in Figure~\ref{figures/accuracy-by-perf}. When the model produces a poor-performing pass list, the quality of the generated code is lower.

\section{Additional Experiments}

In the previous section, we evaluated the performance of an LLM trained to optimize LLVM-IR for code size. In this section, we build additional models to better understand the properties of LLMs for code optimization. All models use the same architecture and parameters as in Section~\ref{section/model}.

\subsection{Abalation of Dataset Size}

We ablate the contribution of dataset size by training two additional models and varying the amount of the training data from 50\% (500k examples) down to 25\% (250k examples) by random dropout. Figure~\ref{figures/ablations} shows progress during the training of the models. For dataset sizes of 50\% and 25\%, the models begin to overfit the training set after around 8B training tokens. Table~\ref{tables/ablations} shows the peak performance of each configuration. With 50\% and 25\% of the training data, downstream performance falls by 21\% and 24\%, respectively.

\subsection{Abalation of Code Optimization Task}%
\label{section/codegen-ablation}

We train the model to generate not just a pass list but also the optimized code resulting from this pass list. One may expect this to degrade model performance -- not only must it learn to predict good pass lists, but also how to produce correctly optimized code, a more difficult task. In fact, we believe this to be crucial to model performance. By forcing LLMs to learn the semantics of LLVM-IR we enable them to make better optimization decisions.

To ablate this we trained a model to generate only pass lists without the corresponding optimized code. We kept the data mix and all other parameters the same. Figure~\ref{figures/ablations} and Table~\ref{tables/ablations} show that without training the model to generate optimized code, downstream performance falls by 16\%.

\subsection{Evaluation of Single Pass Translation}
\label{section/individual-passes}

\begin{figure}
\setcaptiontype{lstlisting}
\begin{minipage}[t]{1\columnwidth}
    \begin{lstlisting}[%
    escapechar=ä%
]
ä\emph{\textcolor{BrickRed}{Optimize the following LLVM-IR using -name-anon-globals:}}ä

ä\textbf{\textcolor{BrickRed}{@0 = private}}ä
ä\textbf{\textcolor{NavyBlue}{@anon.2ef3bda806391c61822366a2a59f2569.0 = private}}ä
ä\textbf{\textcolor{OliveGreen}{@anon.95277a486ffed0b6ba33ab3385b3d7bd.0 = private}}ä
 ä\raisebox{0ex}[0ex][0ex]{\ensuremath{\color{red}\hookrightarrow}}äunnamed_addr constant [14 x i8] c"ä\emph{\textbf{\textcolor{lightgray}{<snip>}}}ä", align 1

define dso_local i32 @f1(i8* %0) {
%2 = call i32 @f2(i8* %0, i8* getelementptr inbounds(
 ä\raisebox{0ex}[0ex][0ex]{\ensuremath{\color{red}\hookrightarrow}}ä[14 x i8], [14 x i8]* 
 ä\raisebox{0ex}[0ex][0ex]{\ensuremath{\color{red}\hookrightarrow}}\textbf{\textcolor{BrickRed}{@0,}}ä
   ä\textbf{\textcolor{NavyBlue}{@anon.2ef3bda806391c61822366a2a59f2569.0,}}ä
   ä\textbf{\textcolor{OliveGreen}{@anon.95277a486ffed0b6ba33ab3385b3d7bd.0,}}ä
  ä\raisebox{0ex}[0ex][0ex]{\ensuremath{\color{red}\hookrightarrow}}äi64 0, i64 0))
 ret i32 %2
}
    \end{lstlisting}
    \subcaption{Failure due to incomplete information. The \lstinline{-name-anon-globals} pass uses the module name to compute a hash. Lacking this, the model hallucinates a random hash.}
    \label{listings/name-anon-globals}
\end{minipage}
\begin{minipage}[t]{1\columnwidth}
\begin{lstlisting}[%
    escapechar=ä,%
]
ä\textcolor{BrickRed}{\emph{Optimize the following LLVM-IR using -instcombine:}}ä

@var_12 = external dso_local global i64, align 8
@var_13 = external dso_local global i32, align 4
@var_14 = external dso_local global i32, align 4

define dso_local void @f1(i64 %arg) {
 ä\textcolor{BrickRed}{\textbf{\%tmp = alloca i64, align 8}}ä
 ä\textcolor{BrickRed}{\textbf{store i64 \%arg, i64* \%tmp, align 8}}ä
 ä\textcolor{BrickRed}{\textbf{\%tmp1 = load i64, i64* \%tmp, align 8}}ä
 ä\textcolor{BrickRed}{\textbf{\%tmp2 = sub i64 0, \%tmp1}}ä
 ä\textcolor{BrickRed}{\textbf{\%tmp3 = sub i64 0, \%tmp2}}ä
 ä\textcolor{BrickRed}{\textbf{store i64 \%tmp3, i64* @var\_12, align 8}}ä
 ä\textbf{\textcolor{NavyBlue}{store i64 \%arg, i64* @var\_12, align 8}}ä
 ä\textbf{\textcolor{OliveGreen}{store i64 0, \qquad i64* @var\_12, align 8}}ä
 store i32 1, i32* @var_13, align 4
 store i32 0, i32* @var_14, align 4
 ret void
}
\end{lstlisting}
\subcaption{Failed data-flow analysis. The model correctly removes redundant instructions but substites the wrong value for a variable. The model-optimized code compiles and has a high BLEU score, but is incorrect.}
\label{listings/instcombine}
\end{minipage}
\caption{Example failures from the pass translation experiment. We combine the model input (red), ground-truth (blue), and model-generated (green) texts into a single unified diff for brevity. Black text is common to all three.}
\end{figure}
\begin{figure}
\centering
\setcaptiontype{lstlisting}
\centerline{\includegraphics[width=.97\columnwidth]{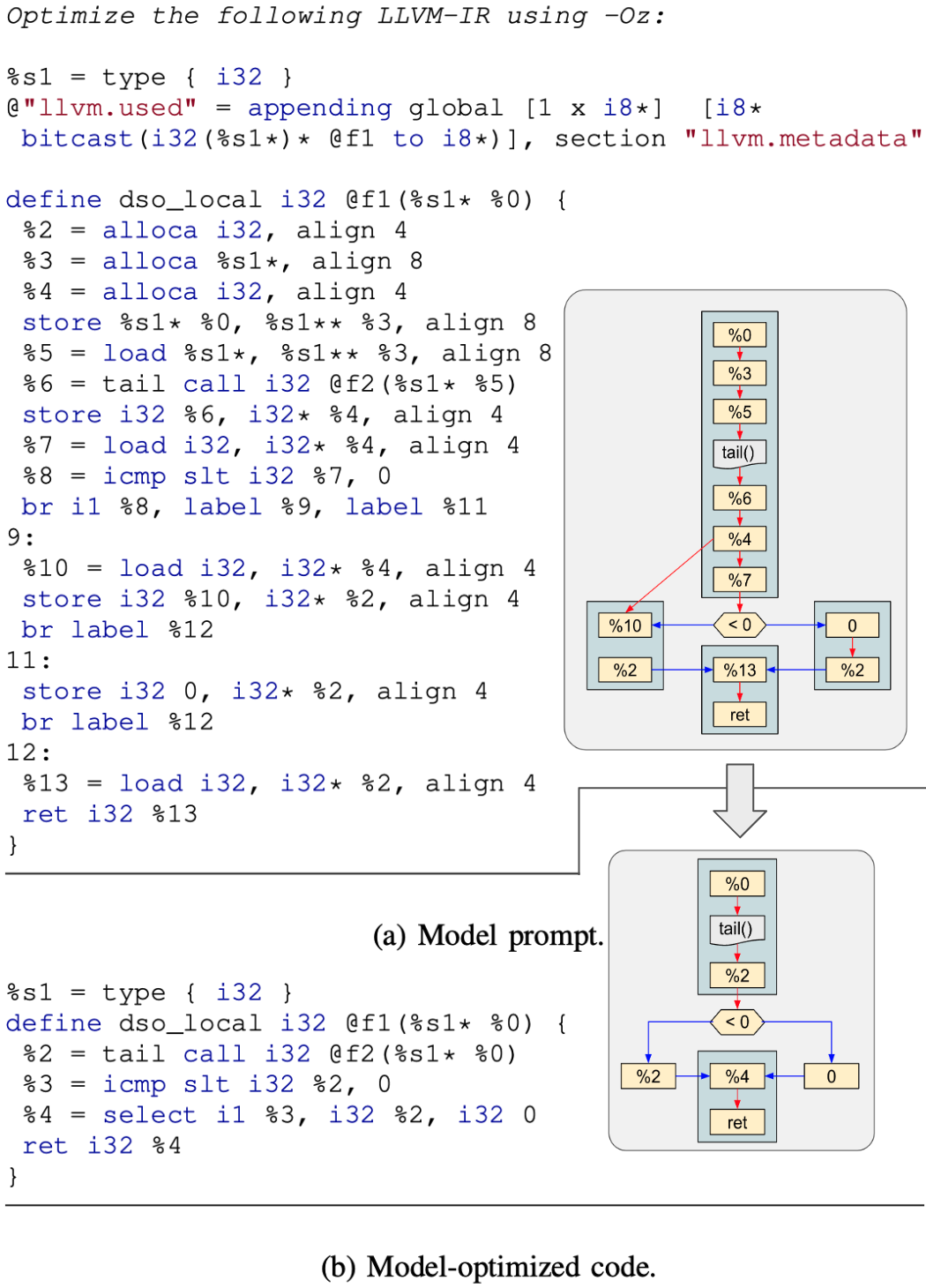}}
\caption{Example of correct generation of optimized IR. The model performed several complex optimizations including control-flow simplification and replacing if-then-else code blocks with instructions.}
\label{listings/success_example}
\end{figure}

In previous sections we trained LLMs to orchestrate optimization passes to produce the best-optimized code. In this section, we evaluate the ability of LLMs to emulate the different optimizations in themselves. For this experiment, the model input is an unoptimized IR and the name of an optimization pass to apply, the output is the IR after applying this pass.

\textbf{Dataset} We generate a new dataset for this task using 60 optimization passes and applying them randomly to the programs from Table~\ref{tables/training-data}. We augment the dataset of unoptimized code with partially optimized code by first running a sequence of randomly selected passes on unoptimized IRs before the desired target pass. We collect 10,000 unique (prompt, answer) examples for each of the 60 passes for a total of 600k examples.

\textbf{Model} We trained a new model from scratch on this pass translation dataset. It reached peak performance after 11B training tokens (74 GPU days).

\textbf{Results} Figure \ref{figures/single-pass-translation} summarizes model performance. The average BLEU score over all passes is 0.846, with exact character-by-character matches 73.7\% of the time and compilable code 82.3\% of the time. We also plot the frequency with which each of the optimizations appears in a model-generated pass list that improved or regressed performance over -Oz in Table~\ref{tables/pass-ordering-results}. We find no correlation between code quality metrics and its frequency in generated pass lists.

As can be seen, many passes are learned near-perfectly while others prove more challenging. Of the passes that perform poorly, some of them hint at simple improvements to the representation while others result from deeper limitations of the model's reasoning. Listing~\ref{listings/name-anon-globals} shows an example from the \lstinline{-name-anon-globals} pass, which is a simple utility pass that renames anonymous global variables using a hash of the module name. Since we do not provide the module name in the prompt, the LLM is forced to hallucinate random values. We will add the module name to prompts to address this.

Listing~\ref{listings/instcombine} shows an example from the \lstinline{-instcombine} pass. This is a complex pass that is implemented in over 4.5k lines of C++ code in LLVM. We see that the model correctly identifies the instructions to combine, but makes an error in data flow analysis and substitutes an incorrect value. This is an important optimization that frequently occurs in pass lists that outperform -Oz. We will explore an active learning approach in which more examples are provided for complex and difficult passes.

Finally, we present an example of correct model optimization in Listing~\ref{listings/success_example}. The example combines several non-trivial code manipulations: register allocation, control flow graph simplification, and instruction combining. We visualize the control- and data-flow graphs to help interpret the changes that the model made. Even on the scale of these small IR functions, we find the sophisticated grasp of LLVM-IR semantics demonstrated by the LLM remarkable. The model has learned to perform these optimizations entirely from examples, without access to the compiler implementation.
\section{Discussion}

We have shown that LLMs can near-perfectly emulate many compiler optimizations and outperform prior approaches, but there are limitations. This section aims to provide a pragmatic discussion of limits and directions for future research.

\subsection{Context Window}

The main limitation of LLMs is the limited sequence length of inputs (context window). In this work we target 2k-token context windows and split IRs into individual functions to maximize the amount of code we can fit into the context window. This is undesirable for a number of reasons. First, it limits the context available to the model when making optimization decisions; second, it prevents intra-function optimization; third, we cannot optimize code that does not fit within the context window. Figure~\ref{figures/improvement-by-size} suggests that larger programs have more interesting optimization opportunities.

Researchers are adopting ever-increasing context windows~\cite{ding2023longnet}, but finite context windows remain a common concern with LLMs. As new techniques for handling long sequences continue to evolve we plan to incorporate them and apply them to code optimization, e.g.\ Code Llama's variant of positional interpolation~\cite{chen2023extending} which is RoPE base period scaling~\cite{llama-code} or recent length extrapolation techniques~\cite{sun2022length}.

\subsection{Math Reasoning and Logic}

Compilers perform lots of arithmetic. Whenever possible expressions are evaluated at compile time to minimize work at runtime and to expose further opportunities for optimization. We see examples of LLMs struggling with this type of reasoning, e.g.\ failed constant folding (Listing~\ref{listings/hard-math}) and failed data-flow analysis (Listing~\ref{listings/instcombine}).

We think that a chain-of-thought approach~\cite{wei2022chain} in which models are taught to decompose complex reasoning problems into incremental steps will prove fruitful. We took the first step in this direction by breaking optimizations down into individual passes in Section~\ref{section/individual-passes}. We also plan to focus training on a curriculum of arithmetic and logic, and train LLMs that use tools to compute intermediate results~\cite{gao2023pal, cobbe2021training}.

\subsection{Inference Speed}

Compilers are fast. It takes two orders of magnitude more time for the model to generate a pass list than it does for the compiler to execute it. While this is much faster than the autotuner it is trained on, it remains an overhead that may prove prohibitive for some applications. That is to say nothing of the difference in compute resources needed to evaluate compiler heuristics vs.\ a 7B-parameter LLM running on multiple GPUs.

In addition to aggressive batching and quantization~\cite{smoothquant}, significant inference speedups can be achieved by specializing the vocabulary to a use case. For example, we can reduce entire subsequences of passes to single vocabulary elements using Byte Pair Encoding so that at inference time fewer tokens need to be generated.


\section{Related Work}

Compiler pass ordering for performance has been exploited for decades~\cite{bodin1998iterative, kisuki_combined_2000, fursin_evaluating_2005}. Over the years there have been several approaches using machine learning~\cite{wang2018machine, ml4sysreview, coreset, autophase, agakov_iterative_2006, ogilvie_minimizing_2017}. The application of machine learning in compilers is not limited to pass order and has been applied to many other problems~\cite{mlgo, Ashouri_2022, hajali2020neurovectorizer, cummins_end2end,phothilimthana2021flexible}. No one has applied LLMs to the problem of pass ordering, we are the first to do so.

\emph{Neural machine translation} is an emerging field that uses language models to transform code from one language to another. Prior examples include compiling C to assembly~\cite{armengol2021learning}, assembly to C~\cite{armengol2023slade,hosseini2022beyond}, and source-to-source transpilation~\cite{transcoder}. In these works code correctness cannot be guaranteed. In our work we use code generation solely as an auxiliary learning task -- correctness is supplied by the compiler.

Language models have found broad adoption for coding tasks, though few operate at the level of compiler IR. Gallagher~et~al.\ 
train a RoBERTA architecture on LLVM-IR for the purpose of code weakness identification~\cite{llvm-weakness-identification} and Transcoder-IR~\cite{transcoder-ir} uses LLVM-IR as a pivot point for source-to-source translation. Neither use LLMs for optimization as we do.

Many language models have been trained on source code including CodeBERT~\cite{codebert}, GraphCodeBERT~\cite{guo2021graphcodebert}, and CodeT5~\cite{codet5} which are trained to perform multiple tasks including code search, code summarization, and documentation generation. LLMs trained on source code have also been used for program fuzzing~\cite{ye_2021,xia2023universal,titanfuzz}, test generation~\cite{schäfer2023adaptive}, and automated program repair~\cite{xia2022training, apr_xia_2023, xia2023conversation}. A large number of useful applications have been explored for language models, however, this is the first work where an LLM is used specifically for optimizing code.

Most LLMs are trained at least partly on code~\cite{chowdhery2022palm,openai2023gpt4,llama1,llama}. Some LLMs are trained similarly to general models but especially target programming languages and can be used for code completion such as Codex~\cite{chen_evaluating_2021} which powers Copilot~\cite{copilot}. The introduction of fill-in-the-middle capabilities is especially useful for real-world code completion use cases and has become common in recent code models such as InCoder~\cite{fried_incoder_2023}, SantaCoder~\cite{allal2023santacoder}, StarCoder~\cite{li_starcoder_2023}, and Code Llama~\cite{llama-code}. Code Llama was also trained to follow instructions and generate code as well as explain its functionalities. 

While the multi-terabyte training corpora for these models contain some assembly, we believe that a focused exploration of the value of LLMs in the domain of compilers will be of value to the community. This paper aims to provide that.
\section{Conclusions}

We present the first steps towards LLMs for code optimization. We construct a model that can predict good optimization strategies for unseen LLVM-IR. Results are promising, though we face challenges in sequence length which limits us to operating over small program fragments, and in arithmetic reasoning which limits the ability of the model to predict the outcome of optimizations. We hope to inspire the research community to push beyond LLMs for simple max-likelihood code generation and into performance-aware code optimization.


\begingroup
\label{bibliography}
\printbibliography

@article{adamw,
  title={{Decoupled Weight Decay Regularization}},
  author={Loshchilov, Ilya and Hutter, Frank},
  journal={arXiv:1711.05101},
  year={2017},
}

@article{armengol2021learning,
  title={{Learning C to x86 Translation: An Experiment in Neural Compilation}},
  author={Armengol-Estap{\'e}, Jordi and O'Boyle, Michael FP},
  journal={arXiv:2108.07639},
  year={2021}
}

@article{armengol2023slade,
  title={{SLaDe: A Portable Small Language Model Decompiler for Optimized Assembler}},
  author={Armengol-Estap{\'e}, Jordi and Woodruff, Jackson and Cummins, Chris and O'Boyle, Michael FP},
  journal={arXiv:2305.12520},
  year={2023}
}

@article{asher2023limits,
      title={{Limits for Learning with Language Models}}, 
      author={Nicholas Asher and Swarnadeep Bhar and Akshay Chaturvedi and Julie Hunter and Soumya Paul},
      year={2023},
      journal={arXiv:2306.12213},
}

@article{Ashouri_2022,
  	year = 2022,
  	author = {Ashouri, Amir H and Elhoushi, Mostafa and Hua, Yuzhe and Wang, Xiang and Manzoor, Muhammad Asif and Chan, Bryan and Gao, Yaoqing},
  	title = {{MLGOPerf: An ML Guided Inliner to Optimize Performance}},
  	journal = {arXiv:2207.08389}
}

@inproceedings{agakov_iterative_2006,
  author={Agakov, F. and Bonilla, E. and Cavazos, J. and Franke, B. and Fursin, G. and O'Boyle, M.F.P. and Thomson, J. and Toussaint, M. and Williams, C.K.I.},
  booktitle={CGO}, 
  title={Using Machine Learning to Focus Iterative Optimization}, 
  year={2006},
}

@inproceedings{ai-soco,
    title={{Overview of the PAN@FIRE 2020 task on the authorship identification of SOurce COde (AI-SOCO)}},
    author={Fadel, Ali and Musleh, Husam and Tuffaha, Ibraheem and Al-Ayyoub, Mahmoud and Jararweh, Yaser and Benkhelifa, Elhadj and Rosso, Paolo},
    booktitle={FIRE},
    year={2020}
}

@article{allal2023santacoder,
    title={{SantaCoder: don't reach for the stars!}},
    author={Loubna Ben Allal and Raymond Li and Denis Kocetkov and Chenghao Mou and Christopher Akiki and Carlos Munoz Ferrandis and Niklas Muennighoff and Mayank Mishra and Alex Gu and Manan Dey and Logesh Kumar Umapathi and Carolyn Jane Anderson and Yangtian Zi and Joel Lamy Poirier and Hailey Schoelkopf and Sergey Troshin and Dmitry Abulkhanov and Manuel Romero and Michael Lappert and Francesco De Toni and Bernardo García del Río and Qian Liu and Shamik Bose and Urvashi Bhattacharyya and Terry Yue Zhuo and Ian Yu and Paulo Villegas and Marco Zocca and Sourab Mangrulkar and David Lansky and Huu Nguyen and Danish Contractor and Luis Villa and Jia Li and Dzmitry Bahdanau and Yacine Jernite and Sean Hughes and Daniel Fried and Arjun Guha and Harm de Vries and Leandro von Werra},
    journal={arXiv:2301.03988},
    year={2023},
}

@inproceedings{apr_xia_2023,
    author = {Xia, Chunqiu Steven and Wei, Yuxiang and Zhang, Lingming},
    title = {{Automated Program Repair in the Era of Large Pre-Trained Language Models}},
    booktitle = {ICSE},
    year={2023}
}

@inproceedings{autophase,
    title={{AutoPhase: Juggling HLS Phase Orderings in Random Forests with Deep Reinforcement Learning}},
    author={Haj-Ali, Ameer and Huang, Qijing and Moses, William and Xiang, John and Wawrzynek, John and Asanovic, Krste and Stoica, Ion},
    booktitle={MLSys},
    year={2020}
}

@inproceedings{bleu,
author = {Papineni, Kishore and Roukos, Salim and Ward, Todd and Zhu, Wei-Jing},
title = {{BLEU: A Method for Automatic Evaluation of Machine Translation}},
booktitle = {ACL},
year = {2002}
}

@inproceedings{bodin1998iterative,
  title={Iterative Compilation in a Non-linear Optimisation Space},
  author={Bodin, Fran{\c{c}}ois and Kisuki, Toru and Knijnenburg, Peter and O'Boyle, Mike and Rohou, Erven},
  booktitle={FDO},
  year={1998}
}

@article{bpe,
  title={{A New Algorithm for Data Compression}},
  author={Gage, Philip},
  journal={C Users Journal},
  volume={12},
  number={2},
  year={1994},
}

@misc{chatgpt,
  author={OpenAI},
  title={{ChatGPT}},
  howpublished = {\url{https://chat.openai.com/}}
}

@article{chen_evaluating_2021,
    title = {{Evaluating Large Language Models Trained on Code}},
    journal = {arXiv:2107.03374},
    author = {Chen, Mark and Tworek, Jerry and Jun, Heewoo and Yuan, Qiming and Pinto, Henrique Ponde de Oliveira and Kaplan, Jared and Edwards, Harri and Burda, Yuri and Joseph, Nicholas and Brockman, Greg and Ray, Alex and Puri, Raul and Krueger, Gretchen and Petrov, Michael and Khlaaf, Heidy and Sastry, Girish and Mishkin, Pamela and Chan, Brooke and Gray, Scott and Ryder, Nick and Pavlov, Mikhail and Power, Alethea and Kaiser, Lukasz and Bavarian, Mohammad and Winter, Clemens and Tillet, Philippe and Such, Felipe Petroski and Cummings, Dave and Plappert, Matthias and Chantzis, Fotios and Barnes, Elizabeth and Herbert-Voss, Ariel and Guss, William Hebgen and Nichol, Alex and Paino, Alex and Tezak, Nikolas and Tang, Jie and Babuschkin, Igor and Balaji, Suchir and Jain, Shantanu and Saunders, William and Hesse, Christopher and Carr, Andrew N. and Leike, Jan and Achiam, Josh and Misra, Vedant and Morikawa, Evan and Radford, Alec and Knight, Matthew and Brundage, Miles and Murati, Mira and Mayer, Katie and Welinder, Peter and {McGrew}, Bob and Amodei, Dario and {McCandlish}, Sam and Sutskever, Ilya and Zaremba, Wojciech},
    year = {2021},
}

@article{chen2023extending,
  title={{Extending Context Window of Large Language Models via Positional Interpolation}},
  author={Chen, Shouyuan and Wong, Sherman and Chen, Liangjian and Tian, Yuandong},
  journal={arXiv:2306.15595},
  year={2023}
}

@article{chowdhery2022palm,
    title={{PaLM: Scaling Language Modeling with Pathways}},
    author={Aakanksha Chowdhery and Sharan Narang and Jacob Devlin and Maarten Bosma and Gaurav Mishra and Adam Roberts and Paul Barham and Hyung Won Chung and Charles Sutton and Sebastian Gehrmann and Parker Schuh and Kensen Shi and Sasha Tsvyashchenko and Joshua Maynez and Abhishek Rao and Parker Barnes and Yi Tay and Noam Shazeer and Vinodkumar Prabhakaran and Emily Reif and Nan Du and Ben Hutchinson and Reiner Pope and James Bradbury and Jacob Austin and Michael Isard and Guy Gur-Ari and Pengcheng Yin and Toju Duke and Anselm Levskaya and Sanjay Ghemawat and Sunipa Dev and Henryk Michalewski and Xavier Garcia and Vedant Misra and Kevin Robinson and Liam Fedus and Denny Zhou and Daphne Ippolito and David Luan and Hyeontaek Lim and Barret Zoph and Alexander Spiridonov and Ryan Sepassi and David Dohan and Shivani Agrawal and Mark Omernick and Andrew M. Dai and Thanumalayan Sankaranarayana Pillai and Marie Pellat and Aitor Lewkowycz and Erica Moreira and Rewon Child and Oleksandr Polozov and Katherine Lee and Zongwei Zhou and Xuezhi Wang and Brennan Saeta and Mark Diaz and Orhan Firat and Michele Catasta and Jason Wei and Kathy Meier-Hellstern and Douglas Eck and Jeff Dean and Slav Petrov and Noah Fiedel},
    year={2022},
    journal={arXiv:2204.02311},
}

@article{cobbe2021training,
  title={{Training Verifiers to Solve Math Word Problems}},
  author={Cobbe, Karl and Kosaraju, Vineet and Bavarian, Mohammad and Chen, Mark and Jun, Heewoo and Kaiser, Lukasz and Plappert, Matthias and Tworek, Jerry and Hilton, Jacob and Nakano, Reiichiro and others},
  journal={arXiv:2110.14168},
  year={2021}
}

@article{codebert,
  title={{CodeBERT: A Pre-trained Model for Programming and Natural Languages}},
  author={Feng, Zhangyin and Guo, Daya and Tang, Duyu and Duan, Nan and Feng, Xiaocheng and Gong, Ming and Shou, Linjun and Qin, Bing and Liu, Ting and Jiang, Daxin and others},
  journal={arXiv:2002.08155},
  year={2020}
}

@article{codesearch-net,
  title={{CodeSearchNet Challenge: Evaluating the State of Semantic Code Search}},
  author={Husain, Hamel and Wu, Ho-Hsiang and Gazit, Tiferet and Allamanis, Miltiadis and Brockschmidt, Marc},
  journal={arXiv:1909.09436},
  year={2019}
}

@article{codet5,
  title={{CodeT5: Identifier-aware Unified Pre-trained Encoder-Decoder Models for Code Understanding and Generation}},
  author={Wang, Yue and Wang, Weishi and Joty, Shafiq and Hoi, Steven CH},
  journal={arXiv:2109.00859},
  year={2021}
}

@inproceedings{compilergym,
  title={{CompilerGym: Robust, Performant Compiler Optimization Environments for AI Research}},
  author={Cummins, Chris and Wasti, Bram and Guo, Jiadong and Cui, Brandon and Ansel, Jason and Gomez, Sahir and Jain, Somya and Liu, Jia and Teytaud, Olivier and Steiner, Benoit and Tian, Yuandong and Leather, Hugh},
  booktitle={CGO},
  year={2022}
}

@misc{copilot,
  title={{Copilot}},
  author={{GitHub}},
  howpublished = {\url{https://copilot.github.com/}}
}

@misc{llvm-canon,
  title={{LLVM Canon}},
  author={Michal Paszkowski},
  howpublished = {\url{https://github.com/michalpaszkowski/LLVM-Canon}}
}

@inproceedings{coreset,
  title={{Learning Compiler Pass Orders using Coreset and Normalized Value Prediction}},
  author={Liang, Youwei and Stone, Kevin and Shameli, Ali and Cummins, Chris and Elhoushi, Mostafa and Guo, Jiadong and Steiner, Benoit and Yang, Xiaomeng and Xie, Pengtao and Leather, Hugh and Tian, Yuandong},
  booktitle={ICML},
  year={2023}
}

@inproceedings{csmith,
  title={{Finding and Understanding Bugs in C Compilers}},
  author={Yang, Xuejun and Chen, Yang and Eide, Eric and Regehr, John},
  booktitle={PLDI},
  year={2011}
}

@inproceedings{cummins_end2end,
  author={Cummins, Chris and Petoumenos, Pavlos and Wang, Zheng and Leather, Hugh},
  booktitle={PACT}, 
  title={{End-to-End Deep Learning of Optimization Heuristics}}, 
  year={2017},
}

@article{difftest,
  title={{Differential Testing for Software}},
  author={McKeeman, William M},
  journal={Digital Technical Journal},
  volume={10},
  number={1},
  year={1998}
}

@article{ding2023longnet,
  title={{LongNet: Scaling Transformers to 1,000,000,000 Tokens}},
  author={Ding, Jiayu and Ma, Shuming and Dong, Li and Zhang, Xingxing and Huang, Shaohan and Wang, Wenhui and Wei, Furu},
  journal={arXiv:2307.02486},
  year={2023}
}

@inproceedings{exebench,
  title={{ExeBench: an ML-scale Dataset of Executable C Functions}},
  author={Armengol-Estap{\'e}, Jordi and Woodruff, Jackson and Brauckmann, Alexander and Magalh{\~a}es, Jos{\'e} Wesley de Souza and O'Boyle, Michael},
  booktitle={MAPS},
  year={2022}
}

@article{fried_incoder_2023,
    title = {{InCoder: A Generative Model for Code Infilling and Synthesis}},
    journal = {arXiv:2204.05999},
    author = {Fried, Daniel and Aghajanyan, Armen and Lin, Jessy and Wang, Sida and Wallace, Eric and Shi, Freda and Zhong, Ruiqi and Yih, Wen-tau and Zettlemoyer, Luke and Lewis, Mike},
    year = {2023},
}

@inproceedings{fursin_evaluating_2005,
	title = {{Evaluating Iterative Compilation}},
	booktitle = {LCPC},
	author = {Fursin, G. G. and O’Boyle, M. F. P. and Knijnenburg, P. M. W.},
	year = {2005},
}

@inproceedings{gao2023pal,
  title={Pal: Program-aided language models},
  author={Gao, Luyu and Madaan, Aman and Zhou, Shuyan and Alon, Uri and Liu, Pengfei and Yang, Yiming and Callan, Jamie and Neubig, Graham},
  booktitle={ICML},
  year={2023},
}

@article{gunasekar2023textbooks,
    title={{Textbooks Are All You Need}},
    author={Suriya Gunasekar and Yi Zhang and Jyoti Aneja and Caio César Teodoro Mendes and Allie Del Giorno and Sivakanth Gopi and Mojan Javaheripi and Piero Kauffmann and Gustavo de Rosa and Olli Saarikivi and Adil Salim and Shital Shah and Harkirat Singh Behl and Xin Wang and Sébastien Bubeck and Ronen Eldan and Adam Tauman Kalai and Yin Tat Lee and Yuanzhi Li},
    year={2023},
    journal={arXiv:2306.11644},
}

@article{guo2021graphcodebert,
    title={{GraphCodeBERT: Pre-training Code Representations with Data Flow}},
    author={Daya Guo and Shuo Ren and Shuai Lu and Zhangyin Feng and Duyu Tang and Shujie Liu and Long Zhou and Nan Duan and Alexey Svyatkovskiy and Shengyu Fu and Michele Tufano and Shao Kun Deng and Colin Clement and Dawn Drain and Neel Sundaresan and Jian Yin and Daxin Jiang and Ming Zhou},
    year={2021},
    journal={arXiv:2009.08366},
}

@inproceedings{hajali2020neurovectorizer,
      title={NeuroVectorizer: End-to-End Vectorization with Deep Reinforcement Learning}, 
      author={Ameer Haj-Ali and Nesreen K. Ahmed and Ted Willke and Sophia Shao and Krste Asanovic and Ion Stoica},
      year={2020},
      booktitle={CGO},
}

@article{hosseini2022beyond,
  title={{Beyond the C: Retargetable Decompilation using Neural Machine Translation}},
  author={Hosseini, Iman and Dolan-Gavitt, Brendan},
  journal={arXiv:2212.08950},
  year={2022}
}

@inproceedings{kisuki_combined_2000,
  author={Kisuki, T. and Knijnenburg, P.M.W. and O'Boyle, M.F.P.},
  booktitle={PACT},
  title={Combined Selection of Tile Sizes and Unroll Factors using Iterative Compilation}, 
  year={2000},
}

@article{li_competition-level_2022,
    title = {Competition-Level Code Generation with {AlphaCode}},
    volume = {378},
    number = {6624},
    journal = {Science},
    author = {Li, Yujia and Choi, David and Chung, Junyoung and Kushman, Nate and Schrittwieser, Julian and Leblond, Rémi and Eccles, Tom and Keeling, James and Gimeno, Felix and Lago, Agustin Dal and Hubert, Thomas and Choy, Peter and d'Autume, Cyprien de Masson and Babuschkin, Igor and Chen, Xinyun and Huang, Po-Sen and Welbl, Johannes and Gowal, Sven and Cherepanov, Alexey and Molloy, James and Mankowitz, Daniel J. and Robson, Esme Sutherland and Kohli, Pushmeet and de Freitas, Nando and Kavukcuoglu, Koray and Vinyals, Oriol},
    year = {2022},
}

@article{li_starcoder_2023,
    title = {{StarCoder}: may the source be with you!},
    journal = {arXiv:2305.06161},
    author = {Li, Raymond and Allal, Loubna Ben and Zi, Yangtian and Muennighoff, Niklas and Kocetkov, Denis and Mou, Chenghao and Marone, Marc and Akiki, Christopher and Li, Jia and Chim, Jenny and Liu, Qian and Zheltonozhskii, Evgenii and Zhuo, Terry Yue and Wang, Thomas and Dehaene, Olivier and Davaadorj, Mishig and Lamy-Poirier, Joel and Monteiro, João and Shliazhko, Oleh and Gontier, Nicolas and Meade, Nicholas and Zebaze, Armel and Yee, Ming-Ho and Umapathi, Logesh Kumar and Zhu, Jian and Lipkin, Benjamin and Oblokulov, Muhtasham and Wang, Zhiruo and Murthy, Rudra and Stillerman, Jason and Patel, Siva Sankalp and Abulkhanov, Dmitry and Zocca, Marco and Dey, Manan and Zhang, Zhihan and Fahmy, Nour and Bhattacharyya, Urvashi and Yu, Wenhao and Singh, Swayam and Luccioni, Sasha and Villegas, Paulo and Kunakov, Maxim and Zhdanov, Fedor and Romero, Manuel and Lee, Tony and Timor, Nadav and Ding, Jennifer and Schlesinger, Claire and Schoelkopf, Hailey and Ebert, Jan and Dao, Tri and Mishra, Mayank and Gu, Alex and Robinson, Jennifer and Anderson, Carolyn Jane and Dolan-Gavitt, Brendan and Contractor, Danish and Reddy, Siva and Fried, Daniel and Bahdanau, Dzmitry and Jernite, Yacine and Ferrandis, Carlos Muñoz and Hughes, Sean and Wolf, Thomas and Guha, Arjun and von Werra, Leandro and de Vries, Harm},
    year = {2023},
}

@article{llama1,
  title={Llama: Open and efficient foundation language models},
  author={Touvron, Hugo and Lavril, Thibaut and Izacard, Gautier and Martinet, Xavier and Lachaux, Marie-Anne and Lacroix, Timoth{\'e}e and Rozi{\`e}re, Baptiste and Goyal, Naman and Hambro, Eric and Azhar, Faisal and others},
  journal={arXiv preprint arXiv:2302.13971},
  year={2023}
}

@article{llama,
  title={{Llama 2: Open Foundation and Fine-Tuned Chat Models}},
  author={Touvron, Hugo and Martin, Louis and Stone, Kevin and Albert, Peter and Almahairi, Amjad and Babaei, Yasmine and Bashlykov, Nikolay and Batra, Soumya and Bhargava, Prajjwal and Bhosale, Shruti and others},
  journal={arXiv:2307.09288},
  year={2023},
}

@article{llama-code,
      title={{Code Llama: Open Foundation Models for Code}}, 
      author={Baptiste Rozière and Jonas Gehring and Fabian Gloeckle and Sten Sootla and Itai Gat and Xiaoqing Ellen Tan and Yossi Adi and Jingyu Liu and Tal Remez and Jérémy Rapin and Artyom Kozhevnikov and Ivan Evtimov and Joanna Bitton and Manish Bhatt and Cristian Canton Ferrer and Aaron Grattafiori and Wenhan Xiong and Alexandre Défossez and Jade Copet and Faisal Azhar and Hugo Touvron and Louis Martin and Nicolas Usunier and Thomas Scialom and Gabriel Synnaeve},
      year={2023},
      journal={arXiv:2308.12950},
}

@inproceedings{llvm,
  title={{LLVM: A Compilation Framework for Lifelong Program Analysis \& Transformation}},
  author={Lattner, Chris and Adve, Vikram},
  booktitle={CGO},
  year={2004},
}

@misc{llvm-weakness-identification,
  title={{LLVM Intermediate Representation for Code Weakness Identification}},
  author={Gallagher, Shannon K and Klieber, William E and Svoboda, David},
  year={2022}
}

@inproceedings{ml4sysreview,
  title={{Machine Learning in Compilers: Past, Present and Future}},
  author={Leather, Hugh and Cummins, Chris},
  booktitle={FDL},
  year={2020},
}

@article{mlgo,
      title={{MLGO: a Machine Learning Guided Compiler Optimizations Framework}}, 
      author={Mircea Trofin and Yundi Qian and Eugene Brevdo and Zinan Lin and Krzysztof Choromanski and David Li},
      year={2021},
      journal={arXiv:2101.04808},
}

@inproceedings{ogilvie_minimizing_2017,
	title = {Minimizing the Cost of Iterative Compilation with Active Learning},
	booktitle = {CGO},
	author = {Ogilvie, William F. and Petoumenos, Pavlos and Wang, Zheng and Leather, Hugh},
	year = {2017},
}

@article{openai2023gpt4,
    title={{GPT-4 Technical Report}},
    author={OpenAI},
    year={2023},
    journal={arXiv:2303.08774},
}

@inproceedings{phothilimthana2021flexible,
  title={{A Flexible Approach to Autotuning Multi-pass Machine Learning Compilers}},
  author={Phothilimthana, Phitchaya Mangpo and Sabne, Amit and Sarda, Nikhil and Murthy, Karthik Srinivasa and Zhou, Yanqi and Angermueller, Christof and Burrows, Mike and Roy, Sudip and Mandke, Ketan and Farahani, Rezsa and others},
  booktitle={PACT},
  year={2021},

}

@inproceedings{poj104,
  title={{Convolutional Neural Networks Over Tree Structures for Programming Language Processing}},
  author={Mou, Lili and Li, Ge and Zhang, Lu and Wang, Tao and Jin, Zhi},
  booktitle={AAAI},
  year={2016}
}

@article{ppo,
  title={{Proximal Policy Optimization Algorithms}},
  author={Schulman, John and Wolski, Filip and Dhariwal, Prafulla and Radford, Alec and Klimov, Oleg},
  journal={arXiv:1707.06347},
  year={2017}
}

@inproceedings{programl,
  title={{ProGraML: A Graph-based Program Representation for Data Flow Analysis and Compiler Optimizations}},
  author={Cummins, Chris and Fisches, Zacharias and Ben-Nun, Tal and Hoefler, Torsten and O'Boyle, Michael and Leather, Hugh},
  booktitle={ICML},
  year={2021}
}

@article{qian2022limitations,
      title={{Limitations of Language Models in Arithmetic and Symbolic Induction}},
      author={Jing Qian and Hong Wang and Zekun Li and Shiyang Li and Xifeng Yan},
      year={2022},
      journal={arXiv:2208.05051},
}

@article{schäfer2023adaptive,
    title={{Adaptive Test Generation Using a Large Language Model}},
    author={Max Schäfer and Sarah Nadi and Aryaz Eghbali and Frank Tip},
    year={2023},
    journal={arXiv:2302.06527},
}

@inproceedings{smoothquant,
  title={{SmoothQuant: Accurate and Efficient Post-Training Quantization for Large Language Models}},
  author={Xiao, Guangxuan and Lin, Ji and Seznec, Mickael and Wu, Hao and Demouth, Julien and Han, Song},
  booktitle={ICML},
  year={2023}
}

@article{sun2022length,
  title={{A Length-Extrapolatable Transformer}},
  author={Sun, Yutao and Dong, Li and Patra, Barun and Ma, Shuming and Huang, Shaohan and Benhaim, Alon and Chaudhary, Vishrav and Song, Xia and Wei, Furu},
  journal={arXiv:2212.10554},
  year={2022}
}

@article{the-pile,
  title={{The Pile: An 800GB Dataset of Diverse Text for Language Modeling}},
  author={Gao, Leo and Biderman, Stella and Black, Sid and Golding, Laurence and Hoppe, Travis and Foster, Charles and Phang, Jason and He, Horace and Thite, Anish and Nabeshima, Noa and others},
  journal={arXiv:2101.00027},
  year={2020}
}

@article{the-stack,
  title={{The Stack: 3TB of Permissively Licensed Source Code}},
  author={Kocetkov, Denis and Li, Raymond and Allal, Loubna Ben and Li, Jia and Mou, Chenghao and Ferrandis, Carlos Mu{\~n}oz and Jernite, Yacine and Mitchell, Margaret and Hughes, Sean and Wolf, Thomas and others},
  journal={arXiv:2211.15533},
  year={2022}
}

@inproceedings{titanfuzz,
    author = {Deng, Yinlin and Xia, Chunqiu Steven and Peng, Haoran and Yang, Chenyuan and Zhang, Lingming},
    title = {Large Language Models Are Zero-Shot Fuzzers: Fuzzing Deep-Learning Libraries via Large Language Models},
    year = {2023},
    booktitle = {ISSTA},
}

@article{transcoder,
  title={{Unsupervised Translation of Programming Languages}},
  author={Lachaux, Marie-Anne and Roziere, Baptiste and Chanussot, Lowik and Lample, Guillaume},
  journal={arXiv:2006.03511},
  year={2020}
}

@article{transcoder-ir,
  title={{Code Translation with Compiler Representations}},
  author={Szafraniec, Marc and Roziere, Baptiste and Charton, Francois and Leather, Hugh and Labatut, Patrick and Synnaeve, Gabriel},
  journal={arXiv:2207.03578},
  year={2022}
}

@article{transformer,
  title={{Attention Is All You Need}},
  author={Vaswani, Ashish and Shazeer, Noam and Parmar, Niki and Uszkoreit, Jakob and Jones, Llion and Gomez, Aidan N and Kaiser, {\L}ukasz and Polosukhin, Illia},
  journal={NeurIPS},
  year={2017}
}

@article{wang2018machine,
    title={{Machine Learning in Compiler Optimisation}}, 
    author={Zheng Wang and Michael O'Boyle},
    year={2018},
    journal={arXiv:1805.03441},
}

@inproceedings{wei2022chain,
  title={Chain-of-thought prompting elicits reasoning in large language models},
  author={Wei, Jason and Wang, Xuezhi and Schuurmans, Dale and Bosma, Maarten and Xia, Fei and Chi, Ed and Le, Quoc V and Zhou, Denny and others},
  booktitle={NeurIPS},
  year={2022}
}

@article{xia2022training,
      title={{Less Training, More Repairing Please: Revisiting Automated Program Repair via Zero-shot Learning}},
      author={Chunqiu Steven Xia and Lingming Zhang},
      year={2022},
      journal={arXiv:2207.08281},
}

@article{xia2023conversation,
      title={{Keep the Conversation Going: Fixing 162 out of 337 bugs for \$0.42 each using ChatGPT}},
      author={Chunqiu Steven Xia and Lingming Zhang},
      year={2023},
      journal={arXiv:2304.00385},
}

@article{xia2023universal,
    title={{Universal Fuzzing via Large Language Models}},
    author={Chunqiu Steven Xia and Matteo Paltenghi and Jia Le Tian and Michael Pradel and Lingming Zhang},
    year={2023},
    journal={arXiv:2308.04748},
}

@inproceedings{yarpgen,
    title={{Random Testing for C and C++ Compilers with YARPGen}},
    author={Livinskii, Vsevolod and Babokin, Dmitry and Regehr, John},
    booktitle={OOPSLA},
    year={2020},
}

@inproceedings{ye_2021,
    year = 2021,
    author = {Guixin Ye and Zhanyong Tang and Shin Hwei Tan and Songfang Huang and Dingyi Fang and Xiaoyang Sun and Lizhong Bian and Haibo Wang and Zheng Wang},
    title = {Automated conformance testing for {JavaScript} engines via deep compiler fuzzing},
    booktitle={PLDI},
}
\balance
\endgroup

\end{document}